\documentclass[twocolumn,aps,prl,superscriptaddress]{revtex4}
\usepackage[utf8]{inputenc}
\usepackage{epsf,graphicx}
\usepackage{multirow}
\usepackage{amsmath,gensymb,amssymb}
\usepackage{lipsum}
\usepackage{dcolumn}
\usepackage{bm}
\usepackage{hyperref}
\usepackage{natbib}
\usepackage{color}

\begin{document}

\title{Direct numerical simulation of acoustic turbulence: Zakharov-Sagdeev spectrum}

\author{E.A. Kochurin}
\email{kochurin@iep.uran.ru}
\affiliation{Institute of Electrophysics, Ural Division, Russian Academy of Sciences, Yekaterinburg, 620016 Russia}
\affiliation{Skolkovo Institute of Science and Technology, 121205, Moscow, Russia}

\author{E.A. Kuznetsov}
\email{kuznetso@itp.ac.ru}
\affiliation{Skolkovo Institute of Science and Technology, 121205, Moscow, Russia}
\affiliation{Lebedev Physical Institute, RAS, 119991 Moscow, Russia}
\affiliation{Landau Institute for Theoretical Physics, RAS, Chernogolovka, 142432 Moscow region, Russia}
\date{\today}

\begin{abstract}
We present the results of direct numerical simulation of three-dimensional acoustic turbulence in medium with weak positive dispersion. It is shown that at the beginning of the long-wavelength region in the turbulence energy distribution in the $k$-space, there  are formed jets in the form of narrow cones. At larger wavenumbers, the cones broaden, and the distribution accordingly tends to isotropic. In this region of wavenumbers, the angle-averaged turbulence spectrum acquires a power-law character, $E(k)\propto k^{-\alpha}$, with the exponent close to $3/2$, which corresponds to the Zakharov-Sagdeev weak acoustic turbulence spectrum.
\end{abstract}

\maketitle

\textbf{Introduction.} As well known (see, for example, \cite{KZ_book, kadomtsev, kadomtsev-kantorovich}), the weak turbulence theory is based on the assumption that the nonlinear interaction of waves is weak in comparison with the  linear wave dispersion, which is determined by the dependence of the second derivative of the eigen frequency  $\omega$ of small amplitude waves relative to the wave vector $\mathbf{k}$. As well known, for both deep water gravitational and capillary waves, the second derivative of $\omega (\mathbf{k})$ is nonzero in the entire range of wavenumbers. Thereby,  initially Gaussian-distributed linear waves with different $\mathbf{k}$ almost remain this property when weak nonlinearity is taken into account. Each wave moving with its own frequency and wave vector experiences the influence of other waves at distances $L$ greater its wavelength, $\propto k^{-1}$. This makes it possible to describe the system of waves statistically in the random phase approximation using kinetic equations for the number of waves (quasiparticles) $n_{k}$ \cite{KZ_book}. In the leading approximation in nonlinearity, the kinetic equations describe either decay processes ($1\rightarrow 2$), when
\begin{equation}
\omega _{k}=\omega _{k_{1}}+\omega
_{k_{2}},\quad\mathbf{k}=\mathbf{k}_{1}+\mathbf{k}_{2},  \label{3wave}
\end{equation}
or in the case of four-wave interaction, the processes of scattering ($2\rightarrow 2$) with the resonance condition
\begin{equation}
\omega _{k}+\omega _{k_{1}}=\omega _{k_{2}}+\omega
_{k_{3}},\quad\mathbf{k}+\mathbf{k}_{1}=\mathbf{k}_{2}+\mathbf{k}_{3}.  \label{4wave}
\end{equation}
From these resonance conditions immediately follows  that for the acoustic type waves, when $\omega _{k}$ depends linearly on the wave vector $\mathbf{k}$, the resonances (\ref{3wave}) and (\ref{4wave}) are satisfied automatically if the wave vectors $\mathbf{k}_{i}$ of all interacting waves are collinear. This means that all resonant waves propagate in the same direction at the same speed.  Strictly speaking, the kinetic equation can not be applied in such a situation. It is necessary to take into account the dispersion of waves, which should ensure the applicability of the weak turbulence theory.  In the dispersion law $\omega _{k}$, one needs to account the next cubic term in the long-wavelength limit. In isotropic media, this expansion is written as
\begin{equation}
\omega _{k}=kc_{s}(1\pm a^{2}k^{2}+...),  \label{disp1}
\end{equation}
where $c_{s}$ is the speed of sound and $a$ is the scale parameter characterizing the wave dispersion. In the case of the sign $+$ in (\ref{disp1}), one says about
positive dispersion,  the different sign  corresponds to  negative
dispersion. Resonance condition (\ref{3wave}) is satisfied only for positive dispersion, and accordingly is not satisfied for a different sign of the dispersion (the latter means that with a weak nonlinearity,  a four-wave process becomes main, when condition (\ref{4wave}) is fulfilled).

In this work, we consider the case of positive dispersion for three-dimensional ($D=3$) acoustic turbulence. As was clarified in \cite{zakh65,zs_70}, the kinetic equation for $D=3$ has a scale-invariant solution of the Kolmogorov type in the long-wavelength limit, independent of the dispersion parameter. This isotropic solution is the Zakharov-Sagdeev turbulence spectrum $E(k)\propto k^{-3/2}$, which describes a direct cascade with a constant energy flux $\varepsilon$ from the long-wave region to the region of large wavenumbers. The fact that the spectrum is independent of the dispersion parameter is related to the integrable singularity in the kinetic equation due to the presence of two $\delta$-functions, i.e., the resonances (\ref{3wave}). In the two-dimensional case, in the region of small $k$, the singularity is non-integrable. As recently found out in\cite{griffin2022energy}, the spectrum in this case is a power-law one, $\propto k^{-1}$, but depends explicitly on the dispersion parameter $a$.

We present the results of direct numerical simulation of three-dimensional acoustic turbulence in a medium with weak positive dispersion. It is shown that jets in the form of narrow cones appear in the energy distribution in the $k$-space in the region close to pumping. At large wavenumbers, the cones broaden, and the distribution tends to isotropic. In this region of wavenumbers, the angle-averaged turbulence spectrum acquires a power-law character, $E(k)\propto k^{-\alpha}$, with an exponent close to $3/2$, which corresponds to the Zakharov-Sagdeev spectrum of weak acoustic turbulence \cite{zs_70}.

\textbf{Basic equations.} Direct numerical simulation of acoustic turbulence was carried out in the framework of the nonlinear string equation for a scalar function $u$ depending on three spatial coordinates $\mathbf{r}=\{x,y,z\}$ and time $ t$:
\begin{equation}
u_{tt}=\Delta u-2a^{2}\Delta ^{2}u+\Delta (u^{2}),  \label{eq0}
\end{equation}
where $a$ is the dispersion parameter introduced above, $\Delta$
is the Laplace operator. Note that this equation in the one-dimensional case refers to equations integrable by the inverse scattering transform \cite{Zakharov1973}. In 3D geometry, this model was first used by Zakharov to study weak acoustic turbulence in \cite{zakh65}. In the linear approximation, the equation (\ref{eq0}) has the dispersion law
\begin{equation}
\omega ^{2}=k^{2}+2a^{2}k^{4},\qquad k=|\mathbf{k}|,  \label{disp}
\end{equation}
coinciding with the Bogolyubov spectrum for the condensate oscillations of a weakly non-ideal Bose gas. The speed of sound $c_{s}$ in this expression is equal to $1$. In the case of weak dispersion $ka\ll 1$, this law transforms into (\ref{disp1}).

The equation (\ref{eq0}) refers to Hamiltonian systems, it can be represented as a system of two equations:
\begin{equation}
u_{t}=\frac{\delta H}{\delta \phi },\qquad \phi _{t}=-\frac{\delta H}{\delta
u},  \label{ham}
\end{equation}
where $\phi $ has the meaning of the hydrodynamic potential, and
the Hamiltonian $H$ is written as
$$H=\frac{1}{2}\int \left[ \left( \nabla \phi \right) ^{2}+u^{2}\right]
d\mathbf{r}+\int a^{2}(\nabla u)^{2}d\mathbf{r}+\frac{1}{3}\int
u^{3}d\mathbf{r}\equiv$$
\begin{equation}\equiv H_{1}+H_{2}+H_{3}.  \label{ham1}
\end{equation}
The Hamiltonian (\ref{ham1}) has three terms. The first $H_{1}$ is the sum of the kinetic and potential energies of linear dispersionless waves. The second term $H_{2}$ is responsible for the dispersive part of the wave energy, and $H_{3}$ describes the nonlinear interaction of waves.

Making the spatial Fourier transform and introducing the normal variables $a_{k}$ and $a_{k}^{\ast }$,
\begin{eqnarray*}
u_{k} &=&\left( \frac{k^{2}}{2\omega _{k}}\right) ^{1/2}(a_{k}+a_{-k}^{\ast
}), \\
\phi_{k} &=&-i\left( \frac{\omega _{k}}{2k^{2}}\right) ^{1/2}(a_{k}-a_{-k}^{\ast
}),
\end{eqnarray*}
equations (\ref{ham}) take the standard form \cite{ZakharovKuznetsov1997}:
\begin{equation}
\frac{\partial a_{k}}{\partial t}=-i\frac{\delta H}{\delta a_{k}^{\ast }},
\label{normal}
\end{equation}
where
\[
H=\int \omega _{k}|a_{k}|^2d\mathbf{k}+\frac{1}{2}\int
V_{k_{1}k_{2}k_{3}}\left( a_{k_{1}}^{\ast
}a_{k_{2}}a_{k_{3}}+\right.\]
\[
\left.+a_{k_{1}}a_{k_{2}}^{\ast }a_{k_{3}}^{\ast }\right)
\delta \left( \mathbf{k}_{1}-\mathbf{k}_{2}-\mathbf{k}_{3}\right)
d\mathbf{k}_{1}d\mathbf{k}_{2}d\mathbf{k}_{3}.
\]
In this Hamiltonian, we left only one resonance term corresponding to decay processes (\ref{3wave}) in the nonlinear term.

In the weak dispersion approximation, we will take into account the dispersion in the quadratic Hamiltonian $H$,
\[
\omega _{k}=k(1+a^{2}k^{2}),
\]
but in the matrix element $V_{k_{1}k_{2}k_{3}}$, it will be neglected:
\[
V_{k_{1}k_{2}k_{3}}=\frac{3}{4\pi ^{3/2}}\left( k_{1}k_{2}k_{3}\right)
^{1/2}.
\]
Hence, the kinetic equation for the pair correlator $n_{k}$ ($\langle a_{k}^{\ast }a_{k_{1}}\rangle =n_{k}\delta (\mathbf{k}-\mathbf {k}_{1})$) in the weak turbulence approximation is written as
\begin{equation}
\frac{\partial n_{k}}{\partial t}=2\pi \int
d\mathbf{k}_{1}d\mathbf{k}_{2}\left( T_{kk_{1}k_{2}}-T_{k_{1}kk_{2}}-T_{k_{2}kk_{1}}\right) ,
\label{kin}
\end{equation}
where
\[T_{kk_{1}k_{2}}=|V_{kk_{1}k_{2}}|^{2}(n_{k_{1}}n_{k_{2}}-n_{k}n_{k_{2}}-\]
\begin{equation}-n_{k}n_{k_{1}})\delta \left(
\mathbf{k}-\mathbf{k}_{1}-\mathbf{k}_{2}\right) \delta \left( \omega
_{k}-\omega _{k_{1}}-\omega _{k_{2}}\right).  \label{delta}
\end{equation}
The turbulence spectrum $E(k)$, i.e., the dependence of the energy on the absolute value of $k$ is found from the solution of this equation after averaging the quantity $\omega _{k}n_{k}k^{2}$ over the entire solid angle $\Omega $:
\[
E(k)=k^{2}\omega _{k}\int n_{k}d\Omega .
\]
For isotropic distributions, obviously $E(k)=4\pi k^{2}\omega _{k}n_{k}$.

As was first noted by Zakharov \cite{zakh65}, in the kinetic equation (\ref{kin}) in the case of isotropic distributions, the dispersion contribution in $\omega _{k}$ can be neglected, despite the presence of the product of two delta functions with respect to frequencies and wave vectors in the collision term giving a singularity in the kinetic equation. This singularity in the kinetic equation turns out to be integrable after averaging over the angles. As a result, the kinetic equation admits a stationary power-law solution: $n_{k}\propto $ $k^{\alpha }$. The exponent $\alpha $ for the Kolmogorov-type spectrum is found using the Zakharov transformations (see \cite{KZ_book}): $\alpha =-11/2$, which corresponds to the Zakharov-Sagdeev spectrum \cite{zs_70}:
\begin{equation}
E(k)=C\varepsilon ^{1/2}k^{-3/2}.  \label{ZS}
\end{equation}
Here $C$ is the Kolmogorov-Zakharov constant, and $\varepsilon $ is the energy dissipation rate per unit volume, which is the energy flux over the spectrum. The power dependence on $\varepsilon $ in the spectrum (\ref{ZS}) with the exponent $1/2$ corresponds to the resonant three-wave interaction.

The existence of the Zakharov-Sagdeev spectrum in the inertial interval as a spectrum of the Kolmogorov type was confirmed in a number of papers \cite{kin1, kin2, kin3} by numerically solving the kinetic equation (\ref{kin}) in the presence of long-wavelength pumping and high-frequency damping. It is important to note that the numerical study of weak acoustic turbulence in the framework of the kinetic equation (\ref{kin}) in the three-dimensional case was carried out only for isotropic distributions. In this paper, we will show that in direct numerical simulation of three-dimensional acoustic turbulence described by the equation (\ref{eq0}) supplemented by damping at large $k$ and pumping in the region of long waves, the structure of the spectra is not isotropic, especially in the region of small $k$.

With account of both pumping and damping terms, the equations (\ref{ham}) are written in the form:
\begin{equation}
u_{t}=-\Delta \phi +\mathcal{F}(\mathbf{k},t)-\gamma _{k}u,  \label{eq1}
\end{equation}
\begin{equation}
\phi _{t}=-u+2a^{2} \Delta u-u^{2},  \label{eq2}
\end{equation}
where the operator $\gamma _{k}$ responsible for dissipation and the forcing term $\mathcal{F}(\mathbf{k},t)$ are given in Fourier space as:
\begin{eqnarray*}
\gamma _{k} &=&0,\quad k\leq k_{d}, \\
\gamma _{k} &=&\gamma _{0},\quad k>k_{d}, \\
\mathcal{F}(\mathbf{k},t) &=&F(k)\cdot \exp [iR(\mathbf{k},t)], \\
F(k) &=&F_{0}\cdot \exp [-(k-k_{1})^{4}/k_{2}^{4}],\quad k\leq k_{2}, \\
F(k) &=&0,\quad k>k_{2}.
\end{eqnarray*}
Here $R(\mathbf{k,}t)$ are random numbers uniformly distributed in the interval $[0,2\pi ]$, $\gamma _{0}$ and $F_{0}$ are constants. The quantity $k_1$ determines the wavenumber on which the maximum pumping amplitude is reached, $k_2$  sets its width, and  $k_3$ corresponds to the scale at which dissipation occurs.

\textbf{Numerical scheme and parameters.} Numerical integration of the system of equations (\ref{eq1}) and (\ref{eq2}) was carried out in a cubic domain $(2\pi )^{3}$ with periodic boundary conditions over all three coordinates. Time integration was performed using explicit scheme  by means of  the Runge-Kutta method of the fourth order accuracy with step $\delta t=2.5\cdot 10^{-3}$. The equations were integrated over spatial coordinates using spectral methods with the total number of harmonics $N^{3}=512^{3}$. To suppress the aliasing effect, we used a filter that nulls higher harmonics with a wavenumber above $k_{a}\geq N/3$. Below we
present results of numerical simulation for the following parameters: $k_{d}=125$, $k_{1}=3$, $k_{2}^{4}=6$, $\gamma_{0}=100$, $a=2.5\cdot 10^{-3}$, $F_{0}=5\cdot 10^{5}$. With this choice of parameters, the inertial interval was more than one decade. The maximum dispersion addition at the end of the inertial interval, $k=k_{d}$, was $(k_{d}a)^{2}\approx 0.1$.

\begin{figure}[t!]
\centering
\includegraphics[width=0.9\linewidth]{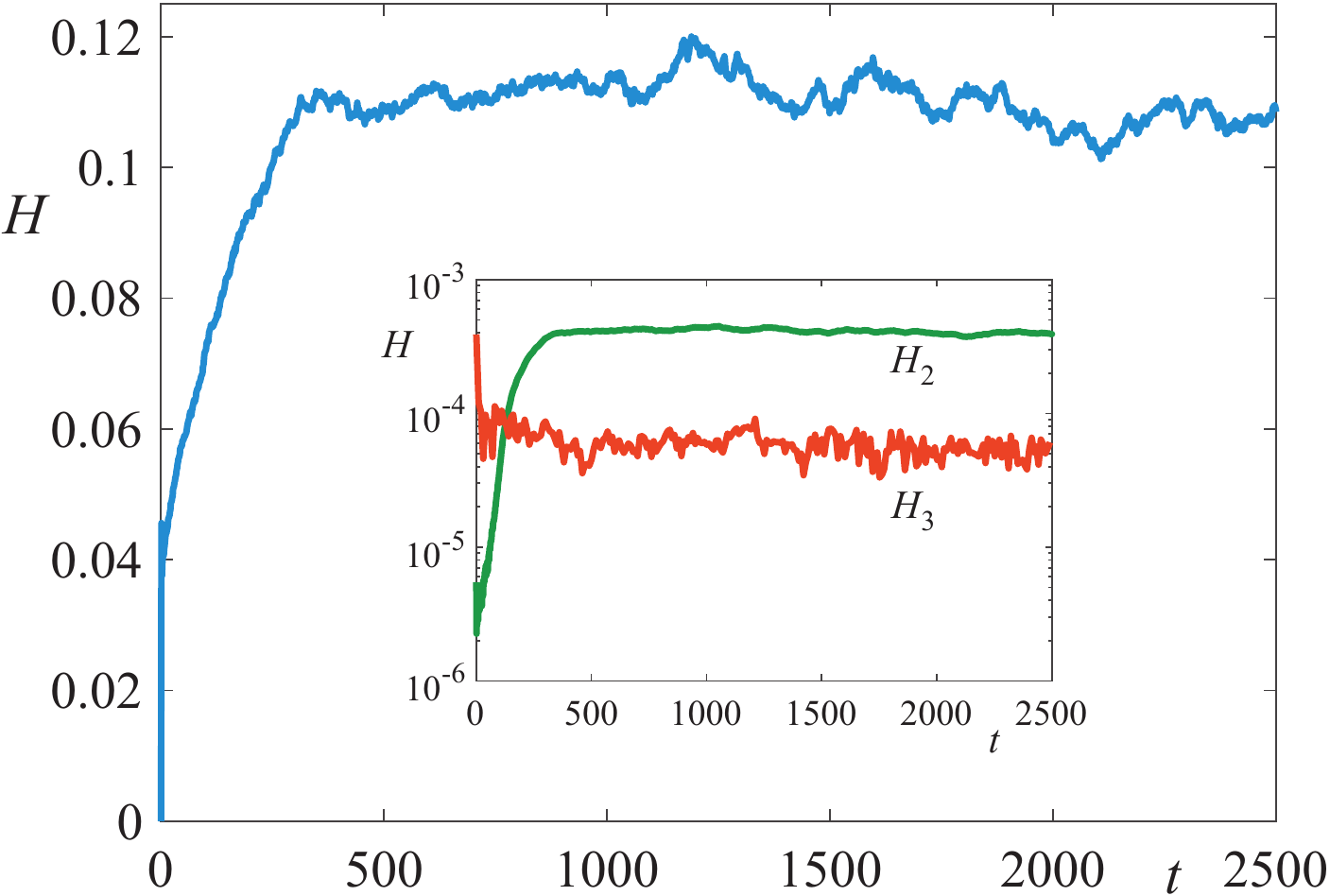}
\caption{(Color online) Total energy of the system (\protect\ref{ham1}) versus time for $a=2.5\cdot 10^{-3}$. The inset shows the time dependencies of the dispersive part of the energy $H_2$ and the nonlinear interaction energy $H_3$.}
\label{fig1}
\end{figure}

\textbf{Simulation results.} In a numerical experiment with the above parameters, we observed a transition to weak turbulence regime. Figure~\ref{fig1} shows how the total energy of the system (\ref{ham1}) evolves. It can be seen the rather quick transition to the regime of quasi-stationary chaotic motion when the effect of pumping in the region of small $k$ is completely compensated by dissipative effects. The inset to Fig.~\ref{fig1} shows the time dependencies of the dispersive part of the energy $H_{2}$ and the nonlinear interaction energy $H_{3}$. Both contributions $H_{2}$ and $H_{3}$ turn out to be small compared to the total energy of the system (respectively, $H_{1}$). The dispersive part of the energy exceeds the energy of the nonlinear interaction by almost an order of magnitude, which indicates on the realization of a weakly nonlinear regime. Thus, the total energy in the inertial interval is approximately equal to $H_{1}\approx\int \epsilon_k d\mathbf{k}$, where $\epsilon_k=k|a_{k}|^2$ is the wave energy density in $k$-space.

\begin{figure}[t!]
\centering
\includegraphics[width=1.0\linewidth]{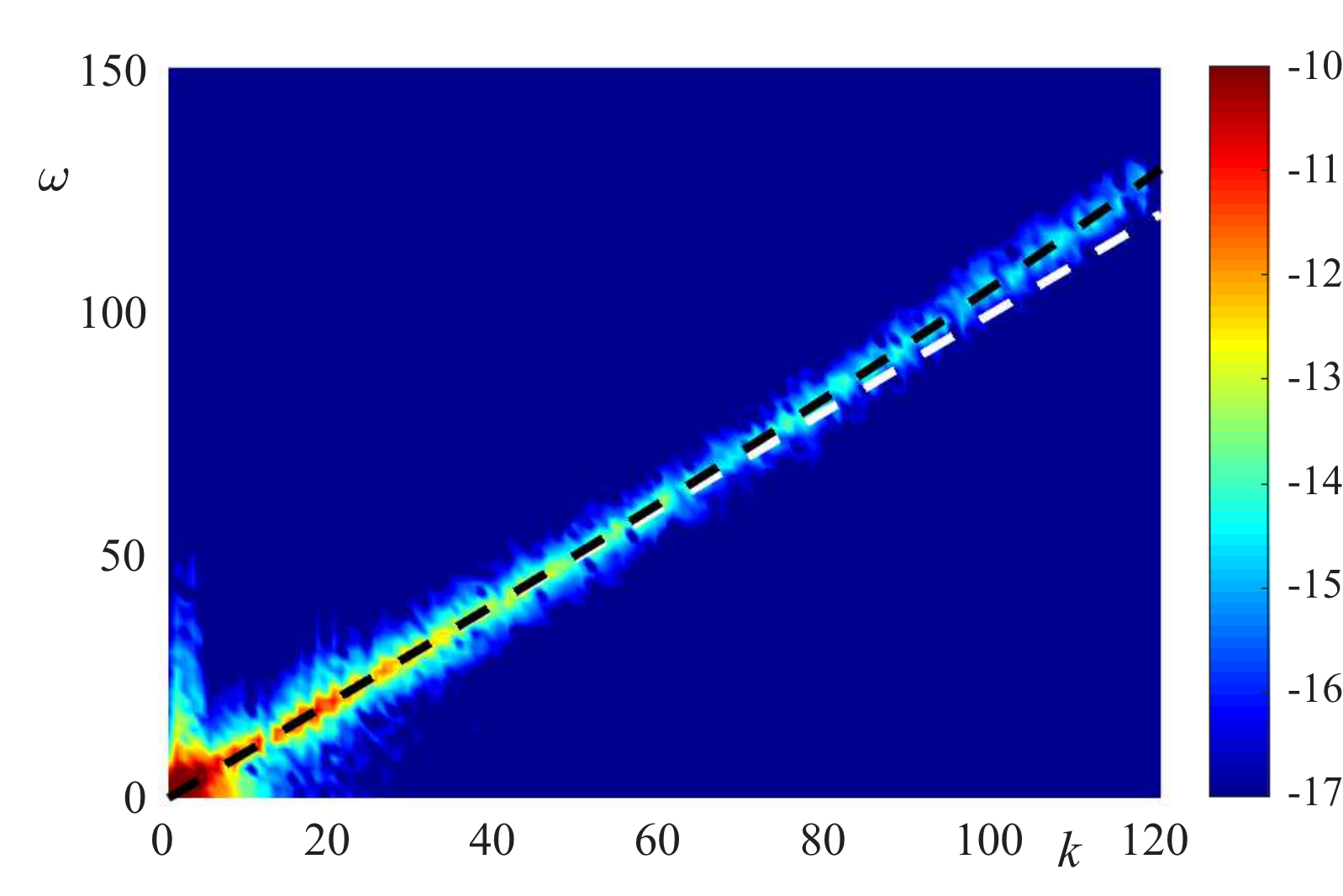}
\caption{ (Color online) The space-time Fourier transform $|u(\mathbf{k},\protect\omega)|^2$ is shown in logarithmic scale. The black dotted line corresponds to the exact value of the dispersion curve (\protect\ref{disp}), the white dotted line corresponds to the non-dispersive wave propagation, $\protect\omega=|\mathbf{k}|$.}
\label{fig2}
\end{figure}

The behavior of the spectrum of space-time Fourier transform of the function $u(\mathbf{r},t)$ shown in Fig.~\ref{fig2} also testifies to the weakly nonlinear character of wave propagation. The figure shows that the wave energy is concentrated along the linear dispersion relation (\ref{disp}). Line broadening is due to nonlinearity. For almost the entire inertial interval, this broadening does not exceed the linear dispersion. For small $k$, the broadening is comparable to the dispersion. For larger $k$, the dispersion exceeds the nonlinear broadening, which agrees with the ratio of the corresponding contributions $H_2$ and $H_3$.

\begin{figure}[t!]
\centering
\includegraphics[width=1.0\linewidth]{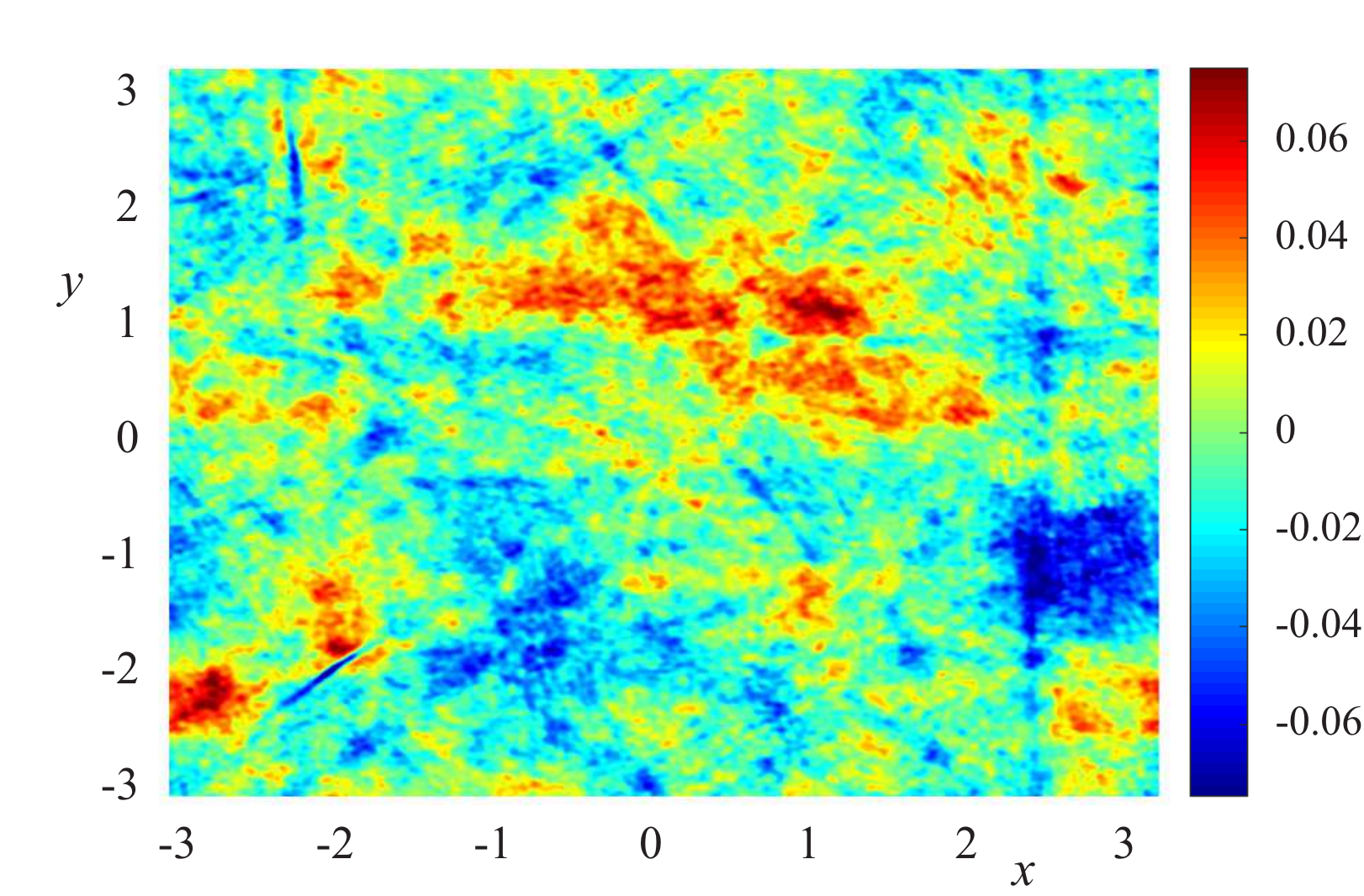}
\caption{(Color online) Section of the function $u(\mathbf{r})$ by $z=0$ plane is shown at the moment $t=2500$ corresponding to the quasi-stationary state.}
\label{fig3}
\end{figure}

The numerical experiment shows that after the system enters the quasi-stationary state, the behavior of $u(\mathbf{r})$ acquires a complex (turbulent) character. In Fig.~\ref{fig3} this behavior demonstrates the dependence of the function $u(\mathbf{r})$ in the $z=0$ plane for the quasi-stationary state at the moment $t=2500$. At the same time, the distribution of the energy density $\epsilon_k$ of turbulent fluctuations in the $k$-space is not isotropic. The anisotropy is especially pronounced in the region of small wavenumbers near the pumping.

\begin{figure}[t!]
\includegraphics[width=1.0\linewidth]{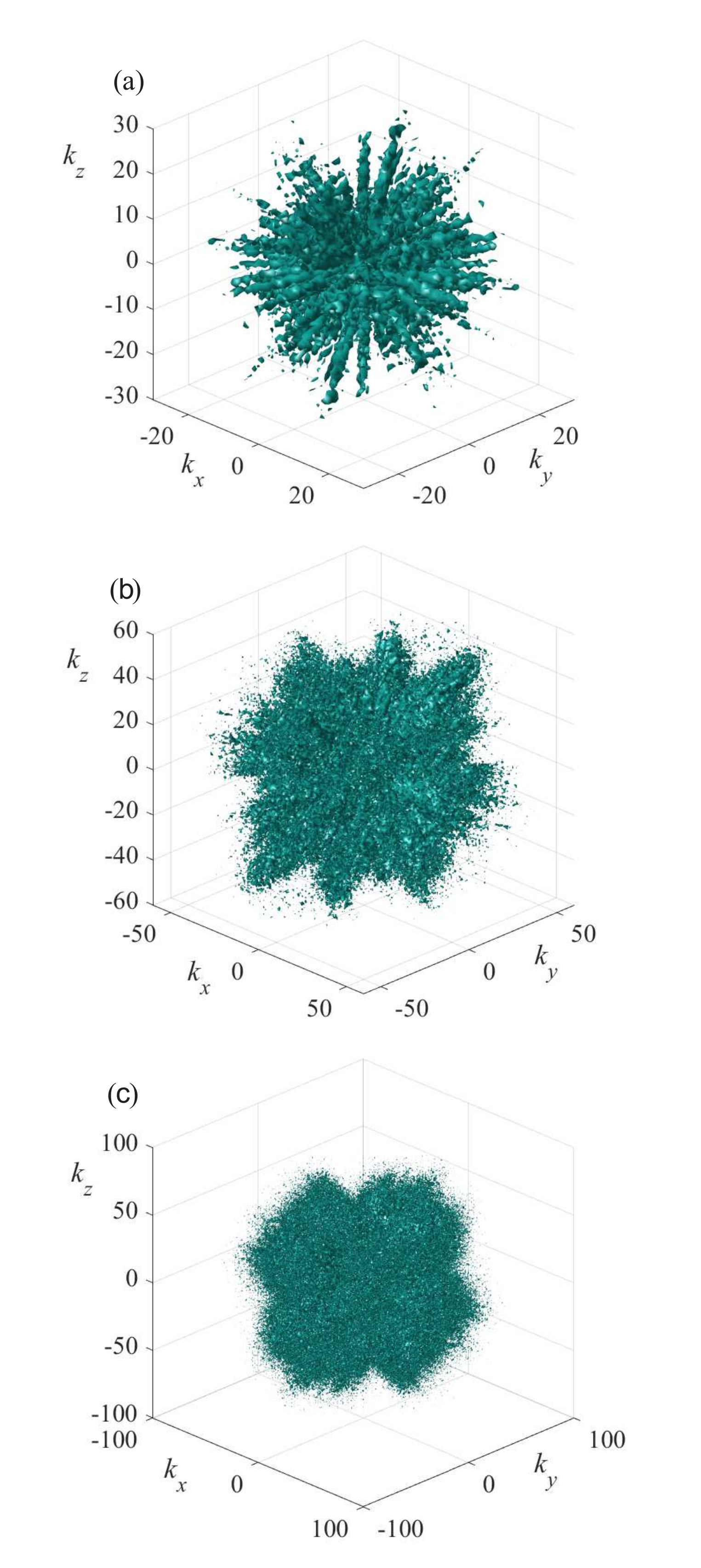}
\caption{(Color online) Isosurfaces of the Fourier spectrum of $|u_\mathbf{k}| =u_0$, (a), (b) and (c) correspond to the values $u_0=5\cdot 10^{-5}$, $0.5\cdot 10^{-5}$ and $0.25\cdot 10^{-5 }$, respectively, $t=2500$.}
\label{fig4}
\end{figure}

\begin{figure}[t!]
\centering
\includegraphics[width=1.0\linewidth]{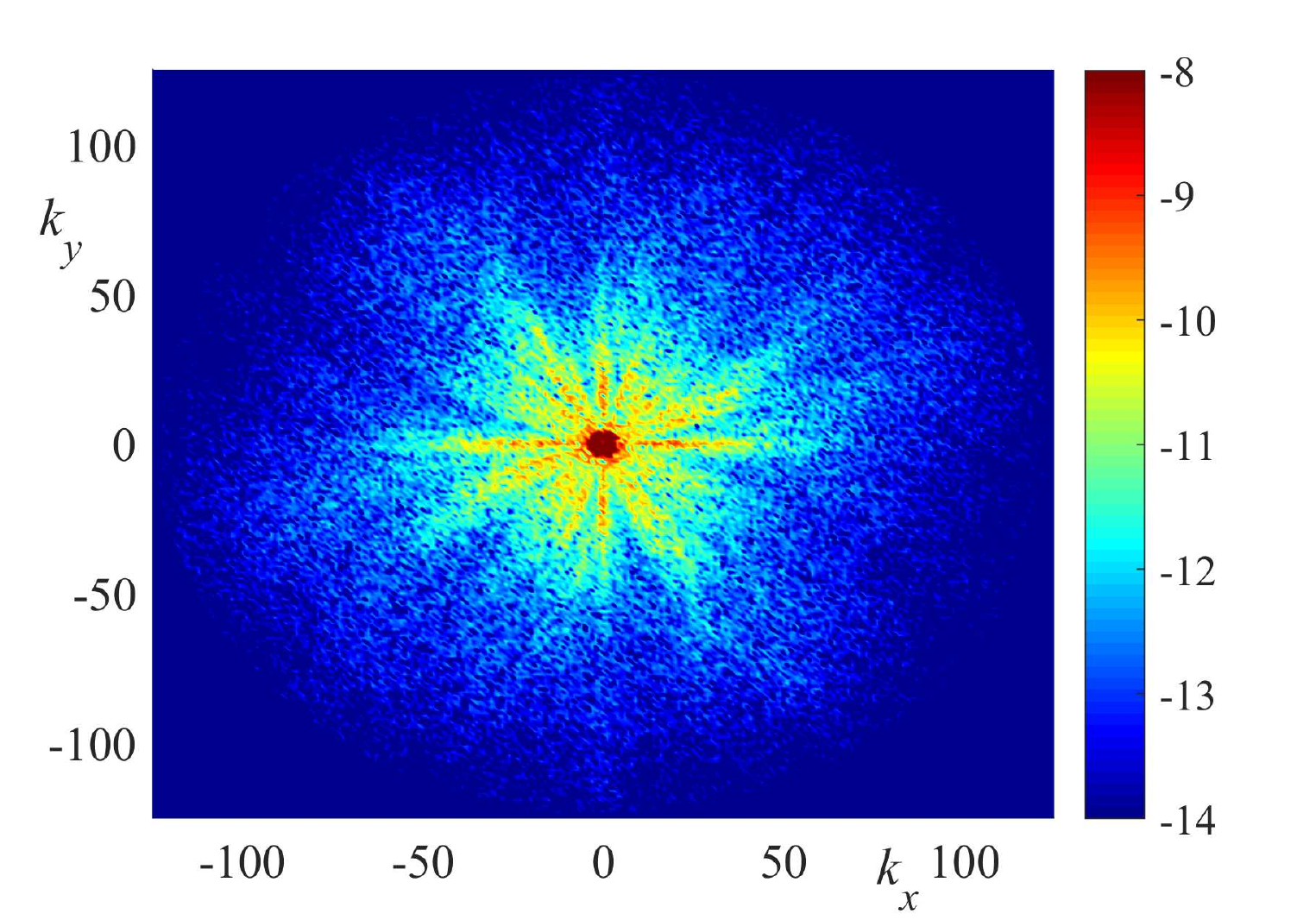}
\caption{(Color online) Fourier spectrum of $|u_{{\bf k}}|\equiv \epsilon_k^{1/2}$ in section $k_z=0$ (logarithmic scale), $t=2500$.}
\label{fig5}
\end{figure}

On fig. ~\ref{fig4} we present three isosurfaces of the function $|u_{\mathbf{k}}|\,(=\epsilon_k^{1/2})$. As seen,  in the region of small wavenumbers, structures with a large number of jets in the form of narrow cones appear in the distribution of turbulent fluctuations. The onset of such structures is the result of resonant wave interactions (\ref{3wave}) at very small $k$ close to the pumping region when dispersion can be neglected. As $k$ increases, the cones broaden and the distribution tends to be isotropic, see Fig.~\ref{fig5}. In this figure, the blue color (at $k\geq 30$) shows a tendency to spectrum isotropization, which is associated with an increase in dispersion with growing $k$ and accordingly with an angular broadening of the resonant surface (\ref{3wave}) by an angle of the order of $ka$.

The generation of jets is associated with two possible causes: linear and non-linear. The first one is the discreteness of the wave vector lattice in the pumping region, $1 \leq k \leq 6$, which inevitably leads to a small anisotropy in the excitation and differences in the growth of perturbations at the initial stage.
The second one is the tendency of the dispersion to zero at small $k$ caused by the equations itself; the three-wave resonance conditions are satisfied for an arbitrary beam. The beams that form the jets have an advantage over other beams. This process, the cooperation of rays into a jet, has a clearly nonlinear character. In particular, this fact follows from the numerical simulation results presented in Fig.~\ref{fig4} and Fig.~\ref{fig5}, for which the contrasts in intensity in the jets and the regions between them are significant: the difference reaches two orders of magnitude. In our opinion, such a jump in intensity can not be explained only by a small anisotropy in pumping, but has a nonlinear origin, possibly due to the collapse of sound waves described by the three-dimensional Kadomtsev-Petviashvili equation (see review \cite{kuznetsov2022}). It should also be noted that numerical experiments show that similar jets appear at the onset of developed hydrodynamic turbulence \cite{AKMS}. However, this question is beyond the scope of this work: new numerical experiments with a high spatial resolution are required.

\begin{figure}[t!]
\centering
\includegraphics[width=0.9\linewidth]{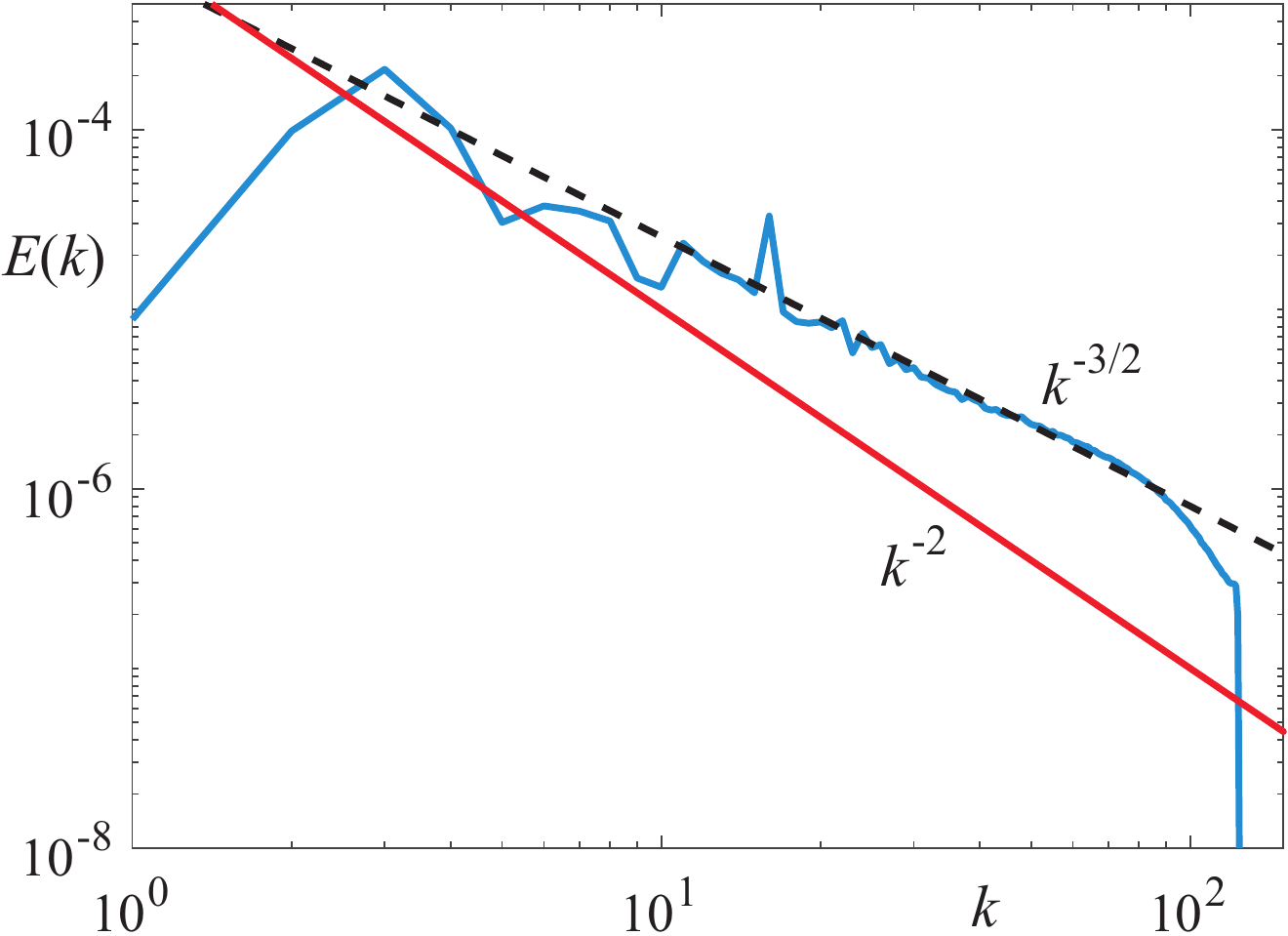}
\caption{(Color online) The turbulence spectrum $E(k)$ measured in the quasi-stationary state, the black dotted line corresponds to the Zakharov-Sagdeev spectrum (\protect\ref{ZS}), the red solid line corresponds to the Kadomtsev-Petviashvili spectrum.}
\label{fig6}
\end{figure}

To find the turbulence spectrum $E(k)$, as noted above, it is necessary to integrate the expression $k^2\epsilon_k$ over the entire solid angle. Figure~\ref{fig6} shows the result of this averaging over the angle for the moment of time corresponding to the quasi-stationary state. It is clearly seen from Fig.~\ref{fig6} that in the regime of quasi-stationary chaotic motion, the spectrum of $E(k)$ acquires a power law behavior. Recall that the inertial interval was $k \in [6, 100]$. As can be seen from Fig.~\ref{fig6}, there are two regions with different behavior of the spectrum in the inertial interval. In the region of large wave numbers, the spectrum of weak acoustic turbulence coincides with the Zakharov-Sagdeev spectrum (\ref{ZS}) with high accuracy, and in the long-wave region, deviations from this spectrum are observed, which, in our opinion, arise due to jets, whose role is significant at small $k$. It should be noted that similar large deviations of an oscillatory nature from the Zakharov-Sagdeev spectrum were observed numerically when simulating the condensate in the framework of the Gross-Pitaevskii equation \cite{proment2012sustained}. In our opinion, these deviations can be related to the anisotropy caused by the presence of jets. No jets were found in the experiment \cite{proment2012sustained}. In our numerical experiment, the spectral energy density $\epsilon_k$ approaches an isotropic distribution in the region of large $k$ as it follows from Fig.~\ref{fig4} and \ref{fig5}. The dispersion $(ka)^2$ changes from $10^{-3}$ to $10^{-1}$ in this region, i.e., remains weak. These two factors contribute to the formation of the Zakharov-Sagdeev spectrum. It should be specially emphasized that the found turbulence spectrum is far from the Kadomtsev-Petviashvili spectrum $E_{KP}\propto k^{-2}$ \cite{KP} related to the dispersionless limit ($a=0$).

\textbf{Conclusion.}
In the present work, direct numerical simulation of three-dimensional acoustic turbulence in a medium with weak positive dispersion is carried out, taking into account the pumping and dissipation of energy. We have established  that the system of nonlinear interacting weakly dispersive waves quickly enough passes into the quasi-stationary chaotic state, which is a developed wave turbulence. In the quasi-stationary regime, the wave field acquires a complex, chaotic character. In the long-wavelength region, close to pumping, we observed in the turbulence spectrum the appearance of narrow jets in the form of cones, which expand upon transition to the short-wave region. The latter leads to the fact that the spectral energy density $\epsilon_k$ tends in the region of large $k$ to an isotropic distribution, for which the dispersion remains weak. In this range of scales, the turbulence spectrum calculated in the stationary state agrees with a high accuracy with the analytical Zakharov-Sagdeev's spectrum of weak acoustic turbulence. It is numerically demonstrated that the criteria of weak turbulence are fully satisfied for this spectrum. Note that the result obtained is the first reliable observation of the spectrum of weak turbulence of acoustic waves in media with positive dispersion in direct three-dimensional numerical simulations.

The authors thank V.E. Zakharov for helpful discussions. This work was financially supported by the Russian Science Foundation (grant no. 19-72-30028).

\end{document}